\documentstyle[prl,multicol,aps]{revtex}

\begin{document}
\setlength{\textwidth}{180mm}
\setlength{\textheight}{240mm}
\setlength{\parskip}{2mm}

\input{epsf.tex}
\title{Two-color multistep cascading and parametric 
       soliton-induced waveguides}

\author{Yuri S. Kivshar$^1$, Andrey A. Sukhorukov$^1$, 
        and Solomon M. Saltiel$^2$}

\address{$^1$ Australian Photonics Cooperative Research Centre,  Research
School of Physical Sciences and Engineering \\  Optical Sciences Center,
Australian National University, Canberra ACT 0200, Australia \\
$^2$ Quantum Electronics Department, Faculty of Physics, University of
Sofia, Sofia 1164, Bulgaria}

\maketitle

\begin{abstract}
We introduce the concept of {\em two-color multistep cascading} for 
vectorial parametric wave mixing in optical media with quadratic 
(second-order or $\chi^{(2)}$) nonlinear response. 
We demonstrate that the multistep cascading allows light-guiding-light 
effects with quadratic spatial solitons. 
With the help of the so-called `almost exact' analytical solutions, we describe
the properties of parametric waveguides created by two-wave quadratic
solitons.
\end{abstract}


\vspace*{-1.0cm}

\begin{multicols}{2}
\narrowtext

Recent progress in the study of cascading effects in optical materials with
quadratic (second-order or $\chi^{(2)}$) nonlinear response has offered 
new opportunities for all-optical processing, optical
communications, and optical solitons \cite{stegeman,chi2_review}. Most of
the studies of cascading effects employ parametric wave mixing
processes with a single phase-matching and, as a result, two-step
cascading. For example, the two-step cascading associated with type I
second-harmonic generation (SHG) includes the generation of the second
harmonic ($\omega + \omega = 2\omega$) followed by reconstruction of the
fundamental wave through the down-conversion frequency mixing (DFM) process
($2\omega - \omega = \omega$). These two processes are governed by one
phase-matched interaction and they differ only in the direction of power
conversion.

The idea to explore more than one simultaneous nearly phase-matched
process, or {\em double-phase-matched (DPM) wave interaction}, became
attractive only recently \cite{assanto,koynov}, 
for the purposes of all-optical 
transistors, enhanced nonlinearity-induced phase shifts, and polarization 
switching. In particular, it was shown \cite{koynov} that multistep 
cascading  can be achieved by two second-order nonlinear cascading processes, 
SHG and  sum-frequency mixing  (SFM), and these two processes 
can also support a novel class of multi-color parametric solitons \cite{ol}. 
The physics involved into the multistep cascading
can be understood by analyzing a chain of parametric processes: 
SHG $(\omega+\omega = 2\omega) \rightarrow$ 
SFM $(\omega + 2 \omega = 3\omega) \rightarrow $ 
DFM $ (3\omega - \omega = 2\omega) \rightarrow $ 
DFM $(2\omega - \omega = \omega)$. 
The main disadvantage of this kind of parametric processes for applications 
is that it requires nonlinear media transparent up to the third harmonic 
frequency.

Then, the important question is: {\em Can we find parametric processes
which involve only two frequencies but allow to get all advantages of
multistep cascading ?} In this Rapid Communication, we answer positively this
question introducing the concept of {\em two-color multistep cascading}. 
We demonstrate a number of unique
features of multistep parametric wave mixing which do not exist for the
conventional two-step cascading. In particular, using one of the
processes of two-color multistep cascading, we show how to introduce
and explore the concept of light guiding light for quadratic spatial
solitons, that has been analyzed earlier
for Kerr-like spatial solitary waves \cite{phys_today} but seemed
impossible for parametric interactions. For the first time to our
knowledge,  we find `almost exact' analytical solutions for two-wave
quadratic solitons and investigate, analytically and numerically, the
properties of parametric waveguides created by quadratic spatial solitons
in $\chi^{(2)}$ nonlinear media.

To introduce more than one parametric process involving only two
frequencies, we consider vectorial interaction of 
waves with different polarization. We denote two orthogonal polarization 
components of the fundamental frequency (FF)
wave ($\omega_1 = \omega$) as A and B, and two orthogonal polarizations 
of the second harmonic (SH) wave ($\omega_2 = 2\omega$),
as S and T. Then, a simple multistep cascading process consists of the
following steps. First, the FF wave A generates the
SH wave S via type I SHG process. Then, by down-conversion
SA-B, the orthogonal FF wave B is generated. At last, the initial
FF wave A is reconstructed by the processes SB-A or AB-S, SA-A. 
Two principal second-order processes AA-S and AB-S correspond to {\em two
different components} of the 
$\chi^{(2)}$
susceptibility tensor, thus introducing additional degrees of freedom into 
the parametric interaction.

\noindent \parbox[l]{8cm}{
\noindent
\begin{table}
\caption{ Possible multistep cascading processes }
\begin{tabular}{cll} 
   & Principal 
   & Equivalent \\ 
\hline
(a) & (AA-S, AB-S)
    & (BB-S, AB-S); (AA-T, AB-T) \\ && (BB-T, AB-T) \\
\hline
(b) & (AA-S, AB-T) 
    & (BB-S, AB-T); (AA-T, AB-S) \\ && (BB-T, AB-S) \\
\hline
(c) & (AA-S, BB-S) & (AA-T, BB-T) \\
\hline
(d) & (AA-S, AA-T) & (BB-S, BB-T) 
\end{tabular}
\label{tab:DPM}
\end{table}
}
Different types of multistep cascading processes are summarized
in Table~\ref{tab:DPM}.
The processes in the row (a) of Table~\ref{tab:DPM} described above and 
the multistep cascading introduced in Ref.~\cite{koynov} are
qualitatively similar, but the latter involves a third-harmonic wave. 

To demonstrate some of the unique properties of the multistep cascading, we 
discuss here how it can be employed for light-guiding-light effects 
in quadratic media. 
For this purpose, we consider 
the principal DPM process (c) (see Table~\ref{tab:DPM}) in the planar 
slab-waveguide geometry. Using the slowly varying envelope approximation with 
the assumption of zero absorption of all interacting waves, we obtain
\begin{equation} \label{eq_1}
 \begin{array}{l}
  {\displaystyle 
    2ik_{1}\frac{\partial A}{\partial z} + \frac{\partial^{2} A}
     {\partial x^{2}} + \chi_{1} S A^{\ast} e^{-i\Delta k_1 z} = 0,}
             \\*[9pt]
  {\displaystyle 
    2ik_{1}\frac{\partial B}{\partial z} + \frac{\partial^{2} B}
     {\partial x^{2}} + \chi_2 S B^{\ast}e^{-i\Delta k_2 z} = 0,}  
             \\*[9pt]
  {\displaystyle 
    4ik_{1}\frac{\partial S}{\partial z} + \frac{\partial^{2} S}
     {\partial x^{2}} + 2 \chi_1 A^2 e^{i\Delta k_1 z} + 2 \chi_2 B^2
                                             e^{i\Delta k_2 z} = 0,} 
  \end{array}
\end{equation}
where $\chi_{1,2} = 2k_1 \sigma_{1,2}$, the nonlinear coupling
coefficients $\sigma_k$ are proportional to the elements of the
second-order susceptibility tensor,
and $\Delta k_1$ and $\Delta k_2$ are the corresponding
wave-vector mismatch parameters.

To simplify the system (\ref{eq_1}), we look for its stationary solutions and
introduce the normalized envelopes $u$, $v$, and $w$ according to the
following relations, 
$A = \gamma_1 u \, \exp ( i\beta z - \frac{i}{2} \Delta k_1 z)$, 
$B = \gamma_2 v \, \exp (i\beta z - \frac{i}{2} \Delta k_2 z)$, and 
$S = \gamma_3 w \, \exp (2i\beta z)$, where $\gamma_1^{-1} = 2\chi_1 x_0^2$, 
$\gamma_2^{-1} = 2x_0^2 (\chi_1 \chi_2)^{1/2}$, and 
$\gamma_3^{-1} = \chi_1 x_0^2$, and the longitudinal and transverse 
coordinates are measured in the units
of $z_0 = (\beta - \Delta k_1/2)^{-1}$ and $x_0 = (z_0/2k_1)^{1/2}$,
respectively. Then, we obtain a system of normalized equations,
\begin{equation}
\label{eq_N}
 \begin{array}{l}
  {\displaystyle 
   i \frac{\partial u}{\partial z} + 
    \frac{\partial^{2}u}{\partial x^{2}}  - u + u^{\ast} w = 0,} 
                     \\*[9pt]
  {\displaystyle 
   i \frac{\partial v}{\partial z} + 
    \frac{\partial^{2} v}{\partial x^{2}} - \alpha_1 v + \chi v^{\ast}w= 0,} 
                     \\*[9pt]
  {\displaystyle 
   2i \frac{\partial w}{\partial z} + 
    \frac{\partial^{2}w}{\partial x^{2}} - \alpha w+\frac{1}{2}(u^2+v^2)= 0,}
  \end{array}
\end{equation}
where $\chi \equiv (\chi_2/\chi_1)$, 
$\alpha_1 = (\beta - \Delta k_2/2)(\beta - \Delta k_1/2)^{-1}$, and 
$\alpha = 4\beta \, (\beta - \Delta k_1/2)^{-1}$.
Equations (\ref{eq_N}) are the fundamental model for describing any type of
multistep cascading processes of the type (c) (see Table~\ref{tab:DPM}).

First of all, we notice that for $v=0$ (or, similarly, $u=0$), the 
dimensionless model (\ref{eq_N}) coincides with the corresponding model for 
the two-step cascading due to type I SHG discussed earlier 
\cite{stegeman,chi2_review}, and its stationary solutions are defined by 
the equations for real $u$ and $w$,
\begin{equation} \label{eq_2}
 \begin{array}{l}
  {\displaystyle 
  \frac{d^2 u}{d x^2}   - u + u w  = 0,} 
          \\*[9pt]
  {\displaystyle 
  \frac{d^2 w}{d x^{2}} - \alpha w + \frac{1}{2} u^2 = 0,}
  \end{array}
\end{equation}
that possess a one-parameter family of two-wave localized solutions
$(u_0, w_0)$ found earlier numerically for any $\alpha \neq 1$, and also
known analytically for $\alpha =1$, 
$u_0(x) = {\left( 3/\sqrt{2} \right)} {\rm sech}^{2} (x/2) = \sqrt{2} w_0(x)$ 
(see Ref. \cite{chi2_review}).

Then, in the small-amplitude approximation,
the equation for real orthogonally polarized FF wave $v$ can
be treated as an eigenvalue problem for an effective waveguide created
by the SH field $w_0(x)$,
\begin{equation}
  \label{eq_eigen}
  \frac{d^2 v}{d x^2} + [\chi \, w_0(x) - \alpha_1] v = 0.
\end{equation}
Therefore, an additional parametric process allows to propagate a probe beam 
of one polarization in {\em an effective waveguide} created
by a two-wave spatial soliton in a quadratic medium with FF component of 
another polarization.
However, this type of waveguide is
different from what has been studied for Kerr-like solitons because it is
{\em coupled parametrically} to the guided modes and, as a result, the 
physical picture of the guided modes is valid, rigorously speaking, only 
in the case of stationary phase-matched beams. As a result, the stability 
of the corresponding waveguide and localized modes of the orthogonal
polarization it guides is a key issue. In particular, the waveguide itself 
(i.e. two-wave parametric soliton) becomes unstable for 
$\alpha < \alpha_{\rm cr} \approx 0.2$ \cite{prl}.

In order to find the guided modes of the parametric waveguide
created by a two-wave quadratic soliton, we have to solve 
Eq.~(\ref{eq_eigen}) where the solution $w_0(x)$ is known numerically only. 
These solutions have been also described by the 
variational method \cite{variational}, but the different types of the 
variational ansatz used do not provide a very good approximation 
for the soliton profile at all $\alpha$.
For our eigenvalue problem
(\ref{eq_eigen}), the function $w_0(x)$ defines 
parameters
of the guided modes and, in order to obtain accurate results, it should be
calculated as close as possible to the exact solutions found numerically.
To resolve this difficulty, below we suggest a novel `almost exact'
solution that {\em would allow to solve analytically many of the 
problems involving quadratic solitons}, including the eigenvalue problem
(\ref{eq_eigen}). 

First, we notice that from the exact result at 
$\alpha =1$ and the asymptotic result for large $\alpha$, 
$w \approx u^2 / {\left( 2\alpha \right)}$, it follows that the SH component 
$w_0(x)$ of Eqs. (\ref{eq_2}) remains almost self-similar for 
$\alpha \geq 1$. Thus, we look for the SH field in 
the form $w_0(x) = w_m \, {\rm sech}^2 (x/p)$, where $w_m$ and $p$ are 
unknown yet parameters. The solution for $u_0(x)$ should be consistent with 
this choice of the shape for SH, and it is defined by the 
first (linear for $u$) equation of the system (\ref{eq_2}). 
Therefore, we can take $u$ in the form of the lowest guided mode, 
$u_0(x) = u_m \, {\rm sech}^p(x/p)$, 
that corresponds to an effective waveguide $w_0(x)$. By matching the 
asymptotics of these trial functions with those defined directly  from 
Eqs.~(\ref{eq_2}) at small and large $x$, we obtain the following solution,
\begin{equation} \label{eq_S}
  u_0(x) = {u_m}{{\rm sech}^{p}(x/p)}, \;\;\; 
  w_0(x) = {w_m}{{\rm sech}^{2}(x/p)},
\end{equation}
\begin{equation} \label{eq_P}
  u_m^2 = \frac{\alpha w_m^2}{\left( w_m -1 \right)}, \;\; 
  \alpha = \frac{4 {\left( w_m-1 \right)}^3}{\left( 2-w_m \right)}, \;\; 
  p= \frac{1}{\left( w_m-1 \right)},
\end{equation}
where all parameters
are functions of $\alpha$ only. It it easy to verify that, for 
$\alpha_{\rm cr} < \alpha < \infty$, the SH amplitude varies 
in the region $1.3 < w_m < 2$, so that all the terms in Eq.~(\ref{eq_P}) 
remain positive. 

\begin{figure}
\setlength{\epsfxsize}{8.0cm}
\centerline{\mbox{\epsffile{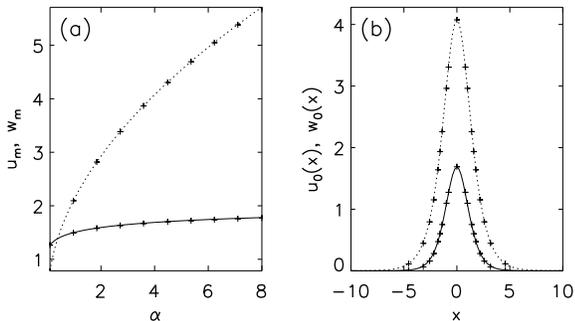}}}
\vspace{2mm}
\caption{ \label{fig:compar}
Comparison between the numerical (continuous curves) and `almost exact'
analytical (crosses) solutions for two-wave (FF -- dotted, SH -- solid)
parametric solitons: 
(a)~Maximum amplitudes;
(b)~Two-wave soliton profile at $\alpha=4$.}
\end{figure}

It is really amazing that the analytical solution (\ref{eq_S}), (\ref{eq_P})  
provides {\em an excellent approximation} for the
profiles of the two-wave parametric solitons found numerically. 
Figures~\ref{fig:compar}(a,b) show a comparison between the maximum 
amplitudes of the FF and SH components and selected soliton 
profiles, respectively.  As a matter of fact, 
the numerical and analytical results on these plots are not distinguishable,
and that is why we show them differently, by continuous curves and crosses. 
For $\alpha < 1$, the SH profile changes, but in the region 
$\alpha > \alpha_{\rm cr}$ the approximate analytical solution 
is still {\em very close to the exact numerical one}: 
a relative error is less than 1\%, for the amplitudes, and 
it does not exceed 3\%, for the power components.  That is why we define 
the analytical solution given by  Eqs.~(\ref{eq_S}), (\ref{eq_P}) as 
`almost exact'.  Details of the derivation, as well as the analysis of 
the case $\alpha <1$, will be presented elsewhere \cite{andrey}.

Now, the eigenvalue problem (\ref{eq_eigen}) can be readily solved 
analytically. The eigenmode cutoff values are defined by the  
parameter $\alpha_1$ that takes one of the discrete values,
$\alpha_1^{(n)} =(s-n)^2/p^2$, where 
$s= - (1/2) + [(1/4) + w_m \chi p^2]^{1/2}$.
Number $n$ stands for the mode order $(n = 0,1, \ldots)$, and the 
localized solutions are possible provided $n <s$. The profiles of the guided
modes can be found analytically in the form
\[
   v_n(x) = V {\rm sech}^{s-n}(x/p) H(-n, 2s-n+1, s-n+1; \zeta),
\]
where $\zeta = \frac{1}{2}[1-\tanh(x/p)]$, $V$ is the mode amplitude,
and $H$ is the hypergeometric function.

\begin{figure}
\setlength{\epsfxsize}{8.0cm}
\centerline{\mbox{\epsffile{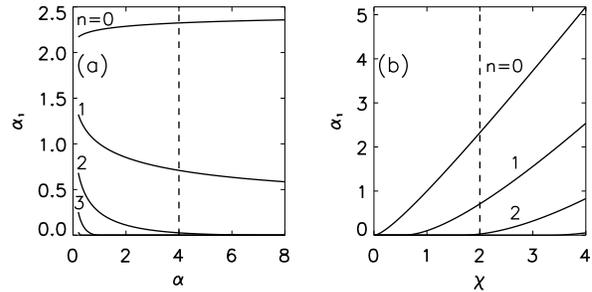}}}
\vspace{2mm}
\caption{ \label{fig:al1}
Cutoff eigenvalues $\alpha_1^{(n)}$ of the guided modes shown
as (a)~functions of $\alpha$ at $\chi=2$, and (b)~functions of $\chi$ at
$\alpha=4$. Dashed lines correspond to the intersection of the plots in
the parameter space $\left( \alpha, \chi \right)$.}
\end{figure}

According to these results, a two-wave parametric soliton creates, in
general, a multi-mode waveguide and
larger number of the guided modes is 
observed for smaller $\alpha$. Figures~\ref{fig:al1}(a,b) show 
the dependence of the mode  cutoff values $\alpha_1^{(n)}$ vs. $\alpha$, at 
fixed $\chi$, and vs. the parameter $\chi$,  at fixed $\alpha$, 
respectively.  For the case $\chi=1$, the dependence has a simple form:
$\alpha_1^{(n)} = [1-n(w_m-1)]^2$.

\begin{figure}
\setlength{\epsfxsize}{8.0cm}
\centerline{\mbox{\epsffile{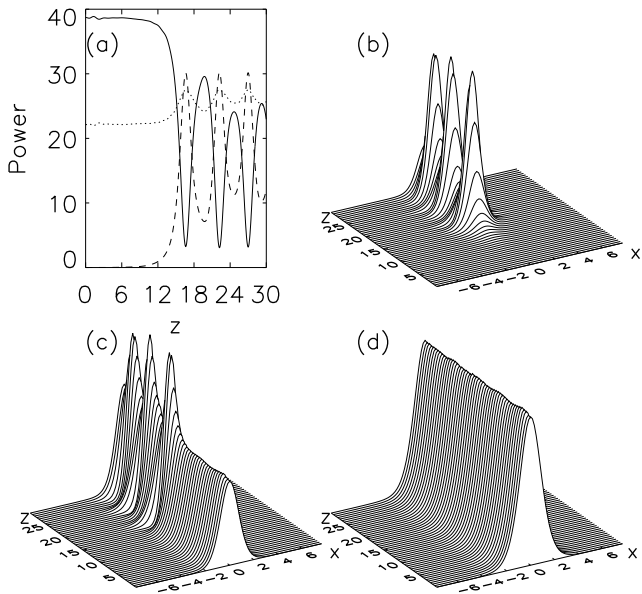}}}
\vspace{2mm}
\caption{ \label{fig:wave_w}
(a)~Change of the normalized power in FF ($u$, solid) and 
SH ($w$, dotted) components, which initially constitute a two-wave 
soliton,
and in the guided mode ($v$, dashed) at $\chi=2$, demonstrating 
amplification of a guided wave.
Evolution of 
the guided wave and effective 
waveguide (SH) is presented in plots (b) and (c), respectively.
(d)~Stationary propagation of a stable fundamental mode ($\chi=1$).
For all the plots $\alpha = 4$, the initial amplitude 
is $v_0 = 0.1$, and $\alpha_1$ corresponds to the bifurcation point.
}
\end{figure}

Because a two-wave soliton creates an induced waveguide parametrically 
coupled to the modes of the orthogonal polarization it guides, the dynamics 
of the guided  modes {\em may differ drastically} from that of conventional 
waveguides based on the Kerr-type nonlinearities. 
Figures~\mbox{\ref{fig:wave_w}(a-d)}  
show two examples of the evolution of guided modes. 
In the first example [see Fig.~\ref{fig:wave_w}(a-c)], 
a weak fundamental mode is amplified via parametric interaction 
with a soliton waveguide, and the mode experiences a strong power 
exchange with the orthogonally polarized FF component through the SH field, 
but with only a weak deformation of the induced waveguide
[see Fig.~\ref{fig:wave_w}(a) -- dotted curve]. This effect 
can be interpreted as a power exchange between two guided modes of
orthogonal polarizations in a waveguide created by the SH field.
In the second example, the propagation is stable 
[see Fig.~\ref{fig:wave_w}(d)]. 

When all the fields in Eq.~(\ref{eq_N}) are not small, i.e. the
small-amplitude approximation is no longer valid,
the profiles of the three-component solitons should be found 
numerically.
However, 
some of the lowest-order states can be calculated
approximately using the 
approach of
the `almost exact' 
solution (\ref{eq_S}),(\ref{eq_P})
described above.
Moreover, a number of the solutions and their families can be 
obtained in {\em an explicit analytical form}. For example, for 
$\alpha_1=1/4$, there exist two
{\em families of three-component solitary waves}
for any $\alpha \geq 1$, that describe
soliton branches starting at the bifurcation points $\alpha_1=\alpha_1^{(1)}$
at \mbox{$\alpha=1$}: (i)~the soliton with a zero-order guided mode 
for $\chi=1/3$:
$u(x) = {\left( 3/\sqrt{2} \right)}\, {\rm sech}^2\left( x/2 \right)$,
$v(x) = c_2\, {\rm sech}\left( x/2 \right)$,
$w(x) = \left( 3/2 \right)\, {\rm sech}^2\left( x/2 \right)$,
and (ii)~the soliton with a first-order guided mode 
for $\chi=1$:
$u(x) = c_1\, {\rm sech}^2\left( x/2 \right)$,
$v(x) = c_2\, {\rm sech}^2\left( x/2 \right) {\rm sinh}\left( x/2 \right)$,
$w(x) = \left( 3/2 \right)\, {\rm sech}^2\left( x/2 \right)$,
where 
$c_2 = \sqrt{3 \left( \alpha-1 \right)}$ and
$c_1 = \sqrt{\left( 9/2 \right) + c_2^2}$.
Some other soliton solutions exist for a specific choice of the 
parameters, e.g. for
$\alpha = \alpha_1 = 4/9$ and $\chi=1$, we find
$u(x) = \left( 4/3 \right)\, {\rm sech}^3\left( x/3 \right)$,
$v(x) = \left( 4/3 \right)\, {\rm sech}^3\left( x/3 \right) 
                             {\rm sinh}\left( x/3 \right)$,
and
$w(x) = \left( 4/3 \right)\, {\rm sech}^2\left( x/3 \right)$.
Stability of these three-wave solitons is a nontrivial issue; 
a rigorous analysis of all such multi-component states is beyond the scope 
of the present Rapid Communication and will be  addressed elsewhere. 

At last, we would like to mention that in the limit of large $\alpha$, when
the coupling to the second harmonic is weak, we can use the cascading
approximation
$w 
   \approx 
  {\left( u^2 + v^2 \right)} / {\left( 2 \alpha \right)}$.
Then, the equations for two orthogonal polarizations of the FF wave reduce 
to a system of two coupled NLS equations, an asymmetric case of \mbox{TE-TM} 
vector spatial solitons well studied in the literature 
(see, e.g., Ref.~\cite{akh} and references therein).

For a practical realization of the DPM processes and the soliton
light-guiding-light effects described above, 
we can suggest two general methods. The first method 
is based on the use of {\em  two commensurable periods} of the 
quasi-phase-matched (QPM) periodic grating.  Indeed, to achieve DPM, we 
can  employ the first-order QPM for one parametric process, and the 
third-order QPM,  for the other parametric process. Taking, as an example, 
the parameters for LiNbO$_3$ and AA-S $(xx-z)$ and BB-S $(zz-z)$ processes
\cite{book}, we find two points for DPM at about 0.89 $\mu$m and 1.25 $\mu$m. 
This means that a  single QPM grating can provide simultaneous phase-matching 
for two parametric  processes. For such a configuration, 
we obtain  $\chi \approx 1.92$ or, interchanging the polarization 
components, $\chi \approx 0.52$.
The second method to achieve the conditions of DPM processes is based on the 
idea of  {\em quasi-periodic QPM grating}. As has been recently shown 
experimentally~\cite{Fib}
and numerically~\cite{us}, 
Fibonacci optical 
superlattices provide an effective way to achieve phase-matching 
at {\em several incommensurable periods} allowing multi-frequency harmonic 
generation in a single structure.

In conclusion, we have introduced the concept of two-color multistep
cascading and demonstrated a possibility of light-guiding-light effects
with parametric waveguides created by two-wave spatial solitons in
quadratic media. We believe our results open a new direction in
research of cascading effects,
and may bring new ideas
into other fields of nonlinear physics where parametric wave interactions
are important.

\end{multicols}
\end{document}